\documentclass[prl,superscriptaddress,showpacs,byrevtex,twocolumn]{revtex4}

\usepackage[]{amssymb}
\usepackage[]{graphicx}
\usepackage{amsmath}
\usepackage{subfigure}
\usepackage{color}
\usepackage{pifont}
\begin{document}

\bibliographystyle{apsrev}

\title{Coherent heavy quasiparticles in CePt$_5$ surface alloy}

\author{M. Klein}
\affiliation{Universit\"at W\"urzburg, Experimentelle Physik VII
\& R\"ontgen Research Center for Complex Materials RCCM, Am Hubland,
D-97074 W\"urzburg, Germany}
\author{A. Nuber}
\affiliation{Universit\"at W\"urzburg, Experimentelle Physik VII
\& R\"ontgen Research Center for Complex Materials RCCM, Am Hubland,
D-97074 W\"urzburg, Germany}
\author{H. Schwab}
\affiliation{Universit\"at W\"urzburg, Experimentelle Physik VII
\& R\"ontgen Research Center for Complex Materials RCCM, Am Hubland,
D-97074 W\"urzburg, Germany}
\author{N. Tobita}
\affiliation{Graduate School of Science, Hiroshima University, Higashi-Hiroshima 739-8526, Japan}
\author{M. Higashiguchi}
\affiliation{Graduate School of Science, Hiroshima University, Higashi-Hiroshima 739-8526, Japan}
\author{J. Jiang}
\affiliation{Graduate School of Science, Hiroshima University, Higashi-Hiroshima 739-8526, Japan}
\author{S. Fukuda}
\affiliation{Graduate School of Science, Hiroshima University, Higashi-Hiroshima 739-8526, Japan}
\author{K. Tanaka}
\affiliation{Graduate School of Science, Hiroshima University, Higashi-Hiroshima 739-8526, Japan}
\author{K. Shimada}
\affiliation{Hiroshima Synchrotron Radiation Center, Hiroshima University, Higashi-Hiroshima 739-0046, Japan}
\author{M. Mulazzi}
\affiliation{Universit\"at W\"urzburg, Experimentelle Physik VII
\& R\"ontgen Research Center for Complex Materials RCCM, Am Hubland, D-97074 W\"urzburg, Germany}
\author{F. F. Assaad}
\affiliation{Universit\"at W\"urzburg, Institut f\"ur Theoretische Physik und Astrophysik, Am Hubland, D-97074 W\"urzburg, Germany}
\author{F. Reinert}\email[corresponding author. Email: ]{reinert@physik.uni-wuerzburg.de}
\affiliation{Universit\"at W\"urzburg, Experimentelle Physik VII
\& R\"ontgen Research Center for Complex Materials RCCM, Am Hubland, D-97074 W\"urzburg, Germany}
\affiliation{Karlsruhe Institute of Technology KIT, Gemeinschaftslabor f\"ur Nanoanalytik, D-76021 Karlsruhe, Germany}

\date{\today}

\begin{abstract}
We report on the results of a high-resolution angle-resolved
photoemission (ARPES) study on the ordered surface alloy CePt$_5$. 
The temperature dependence of the  spectra show the  formation of the
coherent low-energy  heavy-fermion band near the Fermi level.  This experimental data is
supported by a multi-band  model calculation in the framework of the
dynamical mean field theory (DMFT). 

\end{abstract}
\maketitle

The physics of $f$-electrons continues to fascinate since the
interplay between fundamental interactions gives rise to complex phase
diagrams and exotic states of matter \cite{lohneysen07rev}. In Kondo
systems, the two main energy terms determining the microscopical
behavior are the hybridization between localized $f$-states and
conduction electrons and the large Coulomb repulsion between
$f$-electrons at the same lattice site. At temperatures above the
coherence temperature $T^*$
these interactions are purely local and can be explained successfully
by the single-impurity Anderson model (SIAM)
\cite{hewson,GS83a,GS83b,bickers87}. A very powerful tool which allows
to study these local interactions experimentally is photoemission
spectroscopy as it directly probes the spectral function of the system. In the
past the combination of PES and SIAM lead to a deeper understanding of
Kondo systems spectral features like the Kondo resonance (KR) with
additional spin-orbit and crystal-field excitations, including their
temperature dependence \cite{allen05,
ehm07}. Nevertheless, at temperatures $T{\ll}T^*$ the localized heavy
quasiparticles begin to interact with each other and finally form
coherent heavy-fermion bands whose properties cannot be explained by
the SIAM any longer \cite{hewson, fulde}. Up to now experimental
limitations prevented a closer investigation of this coherent
heavy-fermion state by photoemission. In order to observe coherence
effects one usually needs temperatures well below the Kondo
temperature $T_K$ of the system, has to use momentum-resolving probes and an
energy resolution better than $k_BT^*$. A second problem arises from
the surface sensitivity of photoemission techniques, requiring
highly-ordered single-crystalline surfaces to observe dispersing
features. These surfaces can be produced by \textit{in situ\/} cleaving
of layered compounds \cite{allen05, denlinger01, wigger07, vyalikh08,
im08, koitzsch08} or by an \textit{in situ\/} preparation of thin
films of a Ce alloy at the sample surface \cite{garnier97a, danzenbacher05, garnier98,
pillo99}. The second method has the advantage that the surface
quality does not depend on the cleave and the quasi two-dimensional
model systems are comparable to theoretical calculations done
considering the system structure and dimensionality. It is known that the deposition of 4~ML of Ce and subsequent annealing
form a surface alloy that has the same crystal structure as bulk
CePt$_5$, as found by combined studies of scanning tunneling
microscopy, low-energy electron diffraction (LEED) \cite{tang93,baddeley97,berner02,vermang06} and in the reference compound LaPt$_5$ \cite{ramstad99}. This
alloy consists of alternating CePt$_2$ and Pt-\textit{kagome} layers,
whereas the topmost layer is a Pt layer which makes the alloy less
sensitive to surface oxidation. Thus this system is an ideal model
system for studying the effects of the periodic Anderson lattice by
ARPES. Several earlier investigations on CePt$_5$/Pt(111) surface
alloys by photoemission revealed the well known signature of the Kondo
resonance at the Fermi level and even its spin-orbit partner at
$E_{B}\!\approx\!250$~meV \cite{andrews95, garnier97, garnier98,
pillo99}. Nevertheless, those experiments were carried out at either
too high temperatures or limited by instrumental resolution.\\
In this Letter we present high-resolution ARPES measurements and Dynamical Mean Field Theory
(DMFT) calculations on the CePt$_5$/Pt(111) system. We extract the $4f$
spectral weight as a function of the temperature and show for the
first time the transition from a high-temperature phase, in which the system can be
described as a set of isolated impurities, to a low-temperature phase
where coherent heavy fermion quasiparticles become apparent.
The high-resolution ARPES experiments were performed using a Gammadata R4000 analyzer equipped
with a monochromatized discharge lamp using the He II$_\alpha$ emission line
($h\nu\!=\!40.8$~eV, mainly $s$-polarized). The instrumental resolution was determined by
measuring the superconducting gap of a V$_3$Si polycrystal
~\cite{reinert00}, resulting in $\Delta E\!=\!4.3$~meV at the used
spectrometer settings. The resonant
photoemission experiments were carried out at BL1 of the 
HiSOR synchrotron radiation source in Hiroshima, using an SES200
analyzer and linearly polarized light. The total energy resolution
(monochromator and analyzer) was $\Delta E\!=\!30$~meV at $h\nu\!=\!122$~eV.  
Following the method of Ref.~\onlinecite{baddeley97} we prepared
ordered surface alloys by depositing $\approx\!4$~ML of Ce on a Pt(111)
surface. The sample was annealed subsequently for $10$~min at
$\approx\!800$~K. The formation of the CePt$_5$ surface alloy was
evidenced by LEED experiments (not shown) giving a
$(2\times2)R30^{\circ}$-reconstruction, corresponding to a surface alloy with a bulk like CePt$_5$ stoichiometry and geometry \cite{baddeley97}. 

\begin{figure}[ht]
 \begin{center}
 \includegraphics[width=1.0\columnwidth]{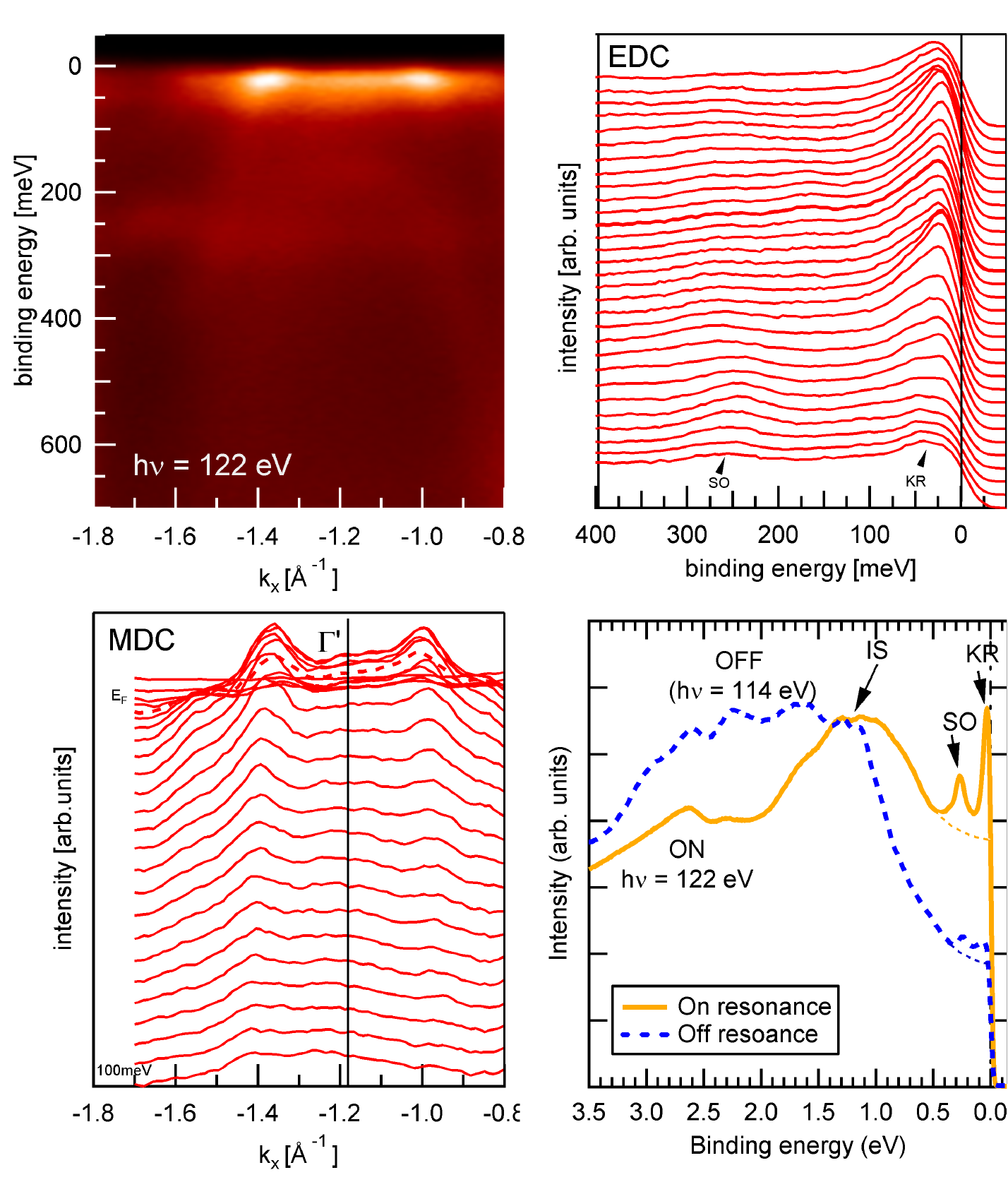}
  \caption{(color online) \textit{Top left}: ARPES data set ($\Delta
E\!=\!16$~meV, $T\!=\!64$~K) around the $\bar{\Gamma}'$-point (second
Brillouin zone) at the $4d$--$4f$ resonance ($h\nu\!=\!122$~eV=). \textit{Top right}: Energy
distribution curves (EDC) taken from top left panel, \textit{bottom left}: Respective momentum distribution
curves (MDC) ($\bar{\Gamma}'$ point indicated by thicker line). The data taken at the Fermi level are
indicated by a thick dashed line. \textit{Bottom right}: Comparison
between the EDC at the $\bar{\Gamma}'$ point measured on- (solid
yellow line) and off-resonance (dashed blue line). KR, SO and IS indicate the Kondo resonance, the spin-orbit partner, and the
ionization structure.} 
 \label{fig:reshigh}
 \end{center}
\end{figure}

To clarify the electronic structure of the CePt$_5$ surface alloy,
Fig.~\ref{fig:reshigh} displays measurements 
at the $4d$--$4f$ resonant excitation energy
($h\nu\!=\!122$~eV) at $T\!=\!64$~K, measured in the second Brillouin
zone around $\bar{\Gamma}'$. The bottom right panel gives the comparison of the
spectrum at the $\bar{\Gamma}'$ point with the off-resonance data
($h\nu\!=\!114$~eV). It shows clearly the main $4f$ features,
namely the KR near the Fermi level $E_F$, the $7/2$
spin-orbit partner (SO) around 0.25~eV, and the ionization peak (IS) at
even higher binding energies.  The KR, with its maximum a few meV
above $E_F$,  reflects the low-energy
excitations of the system and shows a distinct temperature dependence
\cite{bickers87}. ARPES data at the resonance energy (other panels) show again these two
$4f$ structures, without being able to resolve a band dispersion (see
energy distribution curves (EDC, top right panel)),
superimposed by the steep conduction band dispersion, which can be found equivalently in the
reference compound LaPt$_5$ (not shown). The conduction band appears more clearly in the momentum
distribution curves (MDC, bottom left), that represent horizontal cuts
through the color-scaled data set at the top.
In the following we analyse the high-resolution ARPES data in more
detail. Applying a well established method to restore the thermally occupied part of the spectrum up
to an energy of $\approx 5k_BT$ {\em above\/} the Fermi level
\cite{greber97, ehm01b} we divide the ARPES spectra
recorded along the $\bar{\Gamma}\bar{M}$-direction by the Fermi-Dirac function calculated at the experimental temperature.

\begin{figure}[b]
 \begin{center}
 \includegraphics[width=0.9\columnwidth]{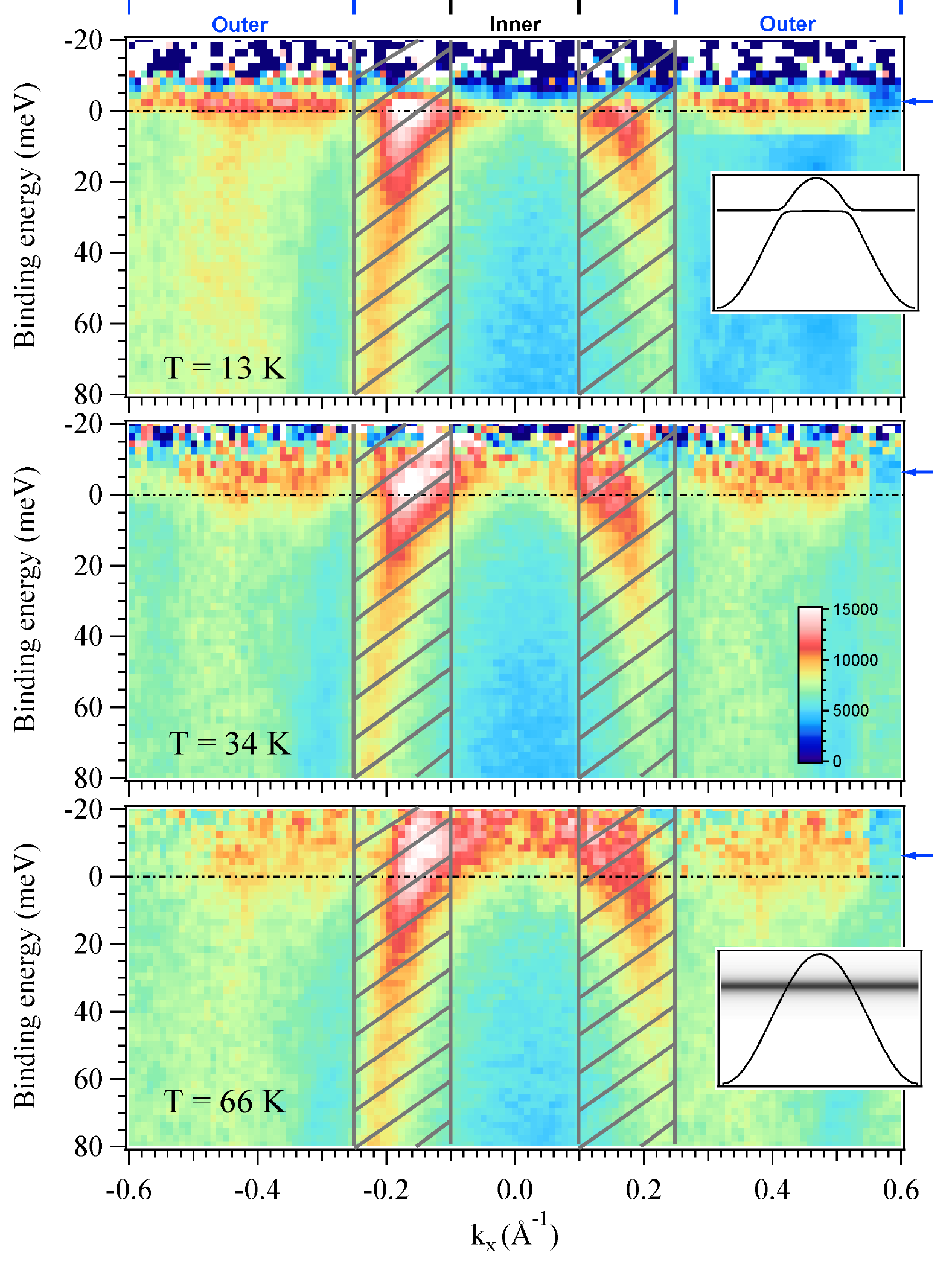}
  \caption{(color online) Temperature dependence of the ARPES data
normalized by the Fermi distribution. The data were acquired as
a function of temperature from the lowest to the highest (top to
bottom: $13$~K, $34$~K and $66$~K). The spectra show a conduction band
(hatched area) and a narrow structure above $E_F$ (indicated by the
blue horizontal arrow), the Kondo resonance. With rising
temperature, the Kondo resonance broadens and shifts away from
$E_F$. The two insets schematically show the transition from the
coherent low-temperature regime to the local regime at high temperatures. 
 \label{fig:temp}} 
 \end{center}
\end{figure}

\begin{figure}[t]
 \begin{center}
  \includegraphics[width=0.9\columnwidth]{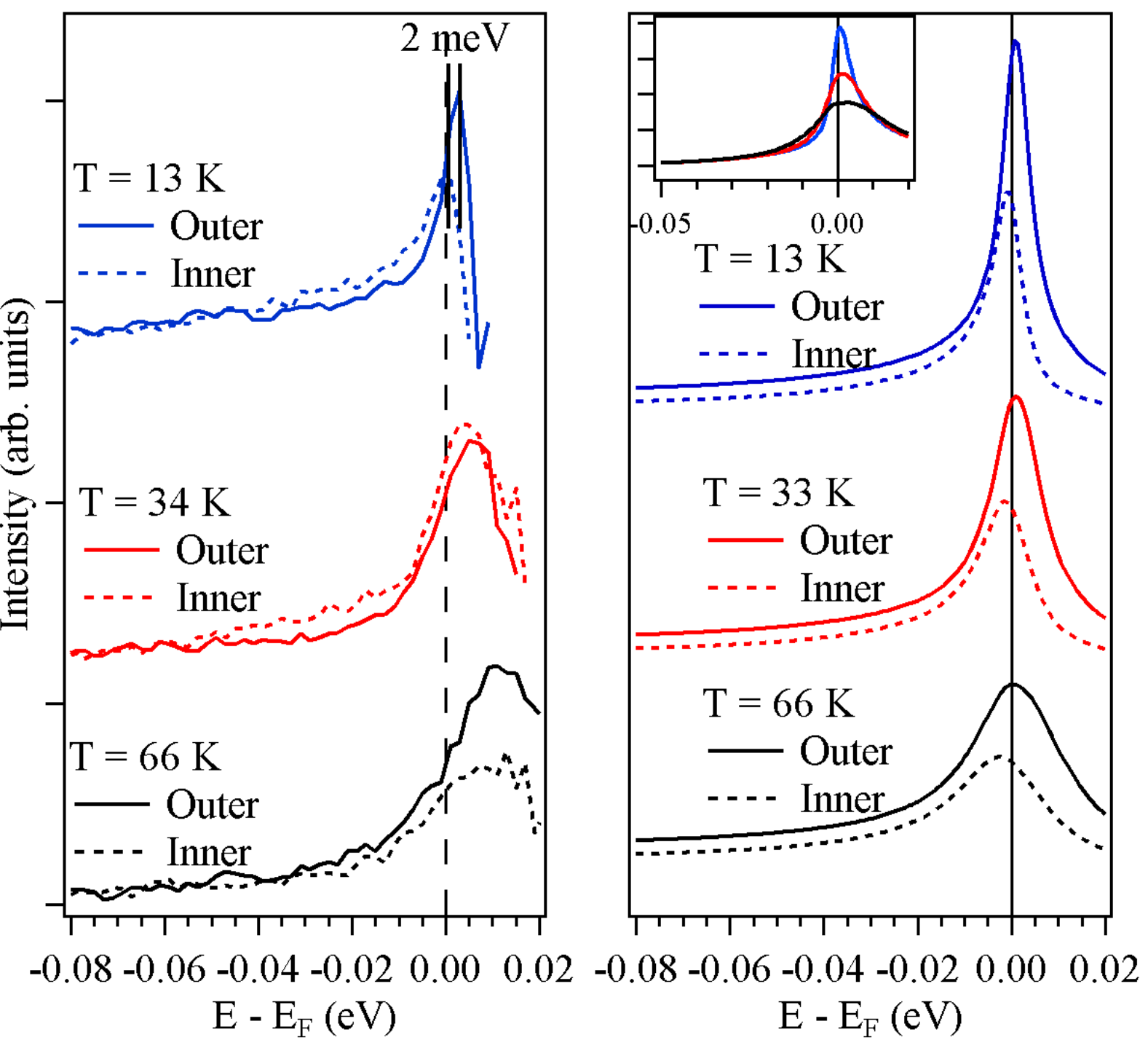}
  \caption{(color online) Left: Integration of the intensity for
$|k|\!>\!0.25$~\AA$^{-1}$ (solid line) and for
$|k|\!<\!0.1$~\AA$^{-1}$ (dashed) at three different temperatures. The
figure shows that at $T\!=\!13$~K the binding energy of the intensity
maxima of the \textit{outer-} and \textit{inner}-~hybridized bands
differ by $\Delta E_{B}\!\approx\!2$~meV. With rising temperature the
difference $\Delta E_{B}$ vanishes. Right: Similar quantity as
obtained from the modeling of the experiment (see text). In the inset
the spectral function integrated over the whole Brillouin zone is
shown. The units of the inset x-scale are electronvolts.} 
 \label{fig:twobands}
 \end{center}
\end{figure}

Fig.~\ref{fig:temp} shows the resulting data for $T\!=\!13$~K, $34$~K and $66$~K. The irregular
black/white pattern above the Fermi level is due to the amplified scatter
in the experimental background. At the lowest temperature ($T\!=\!13$~K) the intensity maximum is
situated right at or slightly above the Fermi level. A few meV below
$E_F$ the conduction band sharply bends towards the
$\bar{\Gamma}$-point. Additionally another very narrow horizontal
band, pinned at the $E_F$, appears at larger $|k|$-values. This
structure seems to be separated from the conduction band around
$\bar{\Gamma}$, as a gapping of the intensity at $|k| =
0.25$~\AA$^{-1}$ suggests. The results are consistent with the
generic mean-field heavy-fermion picture of a flat
\textit{f}-character band hybridizing with the conduction band of the system. This
scenario is sketched in inset to the low temperature data. To estimate
the hybridization gap we plot in Fig.~\ref{fig:twobands} the
integrated intensities taken from two momentum regions:
$|k|\!>\!0.25$~\AA$^{-1}$ and $|k|\!<\!0.1$~\AA$^{-1}$ respectively (as
indicated by the non-hatched area in Fig.~\ref{fig:temp}). The region,
where the intense conduction band lies, is not taken into account by
the integration. The result of the integration is given in
Fig.~\ref{fig:twobands}: It shows that the binding
energies of the intensity maxima differ by $\Delta E_B\!=\!2$~meV. This
shift in energy represents a measure for the size of the hybridization gap due to the
$d-f$ hybridization between the conduction band and the Ce\,$4f$ states. 

It might be noticed that the experimental data shift towards the unoccupied states with
increasing temperature while the maxima in the theoretical spectra
stay constantly at $E_F$. This finding is due to scattering effects in the photoemission
process (both from imperfections and from final state effects). Those
lead to contributions from other regions in
$k$-space, resembling the integrated density of $4f$ states which
shows the observed temperature dependent shift \cite{ehm01}. When our calculated spectral
function is integrated over the whole Brillouin zone, the maximum shifts towards
the unoccupied states as anticipated \cite{bickers87} (inset of
Fig. \ref{fig:twobands}). Since this additional contribution is not
angle dependent, it does not significantly alter the analysis of the
band dispersion.

\begin{figure}[t]
 \begin{center}
 \includegraphics[width=0.9\columnwidth]{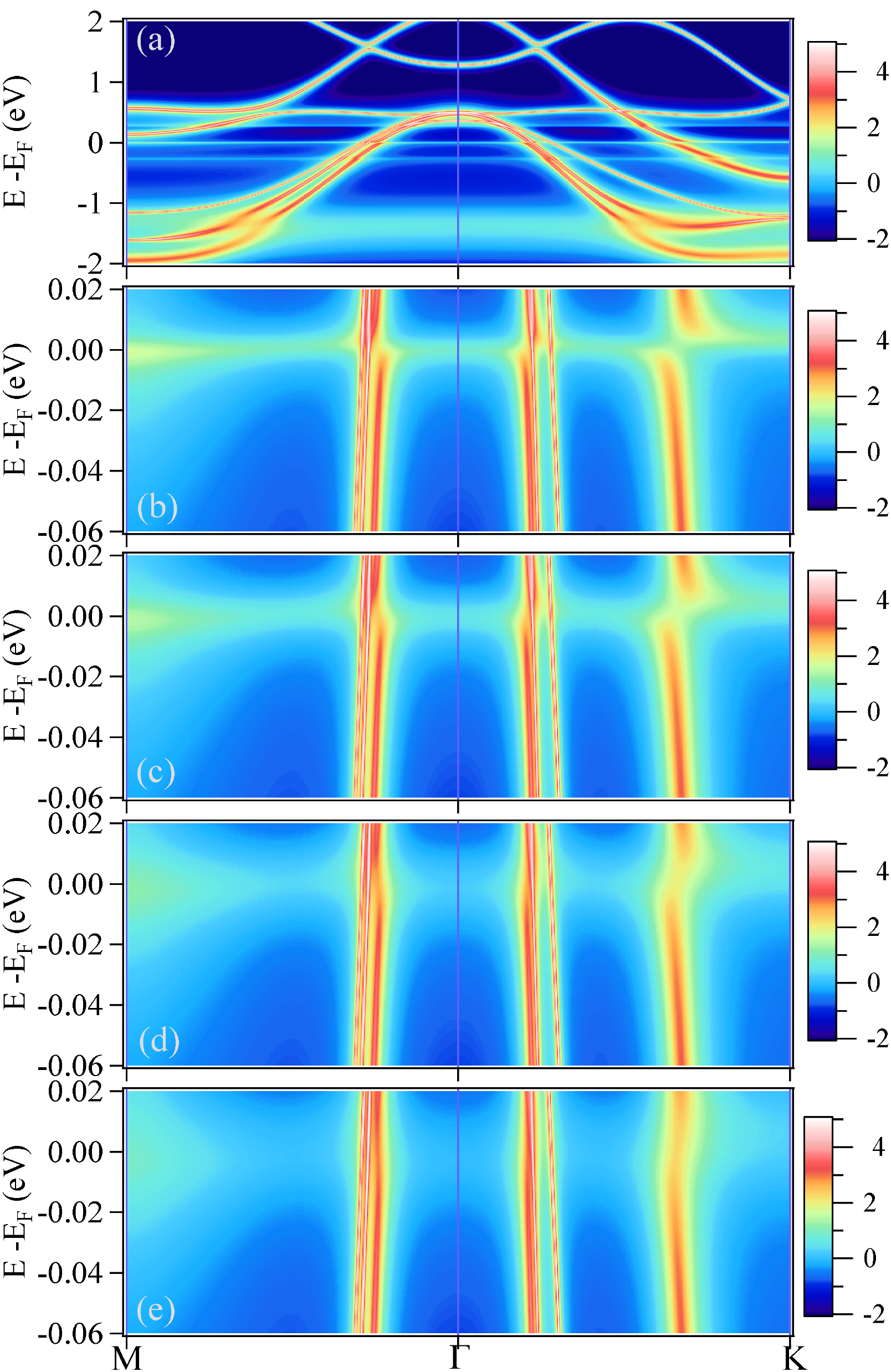}
  \caption{(color online) a) Theoretical total spectral function plotted
at $T\!=\!13$~K vs. energy and momentum in a wide energy
scale. Zoom near the Fermi level with b) $T\!=\!13$~K, c) $T\!=\!33$~K,
d) $T\!=\!66$~K, e) $T\!=\!116$~K. In all the panels the intensities are
plotted in a logarithmic scale indicated in the bars on the right side of each image. As indicated,
the x-axis spans $k$-vectors in the first Brillouin zone from the high
symmetry point $\bar{M}$ over
$\bar{\Gamma}$ to $\bar{K}$.} 
 \label{fig:dmft}
 \end{center}
\end{figure}

The hallmark of the heavy-fermion state lies in its temperature
dependence. As the temperature is raised, the hybridized band
structure gives way to the {\em bare} conduction band thus signaling
the crossover from the heavy Fermion state with large Luttinger volume
to the high temperature phase where the $f$-electrons are
localized and drop out of the Luttinger volume. The two extreme
situations are shown in the insets of Fig.~\ref{fig:temp}. The
temperature dependence of the experimental results is consistent with
this interpretation. The flat band at $|k|\!>\!0.25$~\AA$^{-1}$ broadens
and shifts away from $E_F$. As mentioned above, this temperature
dependence is similar
to the evolution of the Kondo resonance of systems in which the Ce
atoms behave as isolated impurities \cite{bickers87, ehm07}. At the
same time the structure around the $\bar{\Gamma}$-point
($|k|\!<\!0.1$~\AA$^{-1}$) acquires spectral weight and evolves towards
the emergence of a light band crossing the Fermi energy. This growth
in spectral weight is consistent with the closing of the hybridization
gap. In the high temperature limit, i.e. clearly above $T_K$,
and as sketched in the inset, only the conduction band 
disperses (as sketched in the inset). We note that the temperature dependence is consistent with
DMFT calculations for the two-dimensional Kondo lattice
\cite{beach08, suppl}. The starting point for the theoretical modeling of the experiment
presented in Fig.~\ref{fig:dmft} is a tight binding fit of the low
energy features of the bulk reference compound LaPt$_5$. The step to
CePt$_5$ is taken at the model level by including an
$f$-multiplet. The center position of the $f$-multiplet
can be read from Fig.~\ref{fig:reshigh} and is set to $\epsilon_f =
-1.2$~eV. We have included a spin orbit splitting of $300$~meV and set
the Coulomb repulsion to infinity such as to allow solely for $4f^0$
and $4f^1$ configurations. Assuming a $k$-independent constant
hybridization matrix element leaves us with a single free parameter
which we set by fitting at best the temperature dependence of the
experimental data (see Fig.~\ref{fig:twobands}). We solve this
multi-band Anderson model within the DMFT approximation using an non-crossing approximation (NCA)
solver. The details of the approach will be presented elsewhere
\cite{Werner_Assaad_unpublished}. Fig.~\ref{fig:dmft}(a) shows the
total spectral function at low temperature, $T\!=\!13$~K. The sharp
intense features correspond to the LaPt$_5$ bands. At this low
temperature and on this large energy scale we observe distinct rather
dispersionless features which stem from the $f$-electrons: i)
the bare $f$-level at $E-E_f \approx -1.2$~eV ii) the weak spin-orbit
partner of the KR at $E-E_f \approx \pm 0.25$~eV and iii) the KR at $E-E_f
\approx 0$. The temperature evolution of the spectral function, as
given in Figs.~\ref{fig:dmft}(c--f) vs.
energy and momentum, describes nicely the experimental
findings (see Fig.~\ref{fig:temp}). In the considered temperature range, the
emergence of a hybridized band structure, that shows the change of the
heavy quasiparticle
dispersion and - consequently - the change of the Fermi surface
topology, is apparent in this multi-band periodic
Anderson model.

In conclusion we demonstrated that on high quality CePt$_5$ films
grown on Pt(111), high-resolution ARPES experiments allow to extract the
temperature dependence of the $4f$ spectral weight.  
At high temperatures we observe the well known Kondo resonance with a less intense spin-orbit
feature, superimposed by the conduction bands. At low temperatures a
remarkable change in the band dispersion appears: in the vicinity
of the Fermi level the Kondo resonance evolves into hybridized heavy quasiparticle
bands thus signaling, at the single-particle level, the
onset of the coherent heavy-fermion Fermi-liquid state. A comparison with DMFT+NCA calculations of a multi-band
Anderson model which takes into account the band structure of the
reference compound, LaPt$_5$, supports these conclusions and allows a
detailed analysis of the microscopic interactions.
 
\acknowledgments{This work was supported by the JSPS and the Deutsche
Forschungsgemeinschaft within the Forschergruppe 1162. We'd like to
thank Hans Kroha for helpful discussions. The synchrotron-radiation experiments were
done under the approval of HSRC Proposal No. 07-A-42. We acknowledge the FOR1162 DFG project for financial support.}

\end{document}